\documentclass{emulateapj}

\usepackage{amssymb}
\usepackage{amsmath}
\usepackage{natbib}
\usepackage{txfonts}
\usepackage[colorlinks=true,citecolor=blue,linkcolor=blue]{hyperref}
\usepackage{microtype}
\usepackage{color}
\usepackage{graphicx}
\usepackage{epstopdf}

\def\mnras{MNRAS}
\def\apj{ApJ}
\def\apjl{ApJL}
\def\apjs{ApJS}

\def\aap{A\&A}

\def\ufabc{1}
\def\soton{2}

\shorttitle{General relativistic surface degrees of freedom in perturbed hybrid stars}
\shortauthors{Pereira and Lugones}

\begin{document}

\title{General relativistic surface degrees of freedom in perturbed hybrid stars}

\author{Jonas P.~Pereira\altaffilmark{\ufabc,\soton},  and Germ\'an Lugones\altaffilmark{\ufabc}}

\altaffiltext{\ufabc}{Universidade Federal do ABC, Centro de Ci\^encias Naturais e Humanas, Avenida dos Estados, 5001- Bang\'u, CEP 09210-170, Santo Andr\'e, SP, Brazil}

\altaffiltext{\soton}{Mathematical Sciences and STAG Research Centre,
University of Southampton, Southampton, SO17 1BJ, United Kingdom}

\begin{abstract}
We study how the nature of a hybrid system (perfect fluid, solid or a mixture of them) could be related to the induction of general relativistic surface degrees of freedom on phase-splitting surfaces upon perturbation of its phases. We work in the scope of phase conversions in the vicinity of sharp phase transition surfaces whose timescales are either much smaller (rapid conversions) or larger (slow conversions) than the ones of the perturbations ($\omega^{-1}$, where $\omega$ is a characteristic frequency of oscillation of the star). In this first approach, perturbations are assumed to be purely radial. We show that surface degrees of freedom could emerge when either the core or the crust of a hybrid star is solid and phase conversions close to a phase-splitting surface are rapid. We also show how this would change the usual stability rule for solid hybrid stars, namely $\partial M_0/\partial \rho_c\geq 0$, where $M_0$ is the total mass to the background hybrid star and $\rho_c$ its central density. Further consequences of our analysis for asteroseismology are also briefly discussed.
\end{abstract}


\altaffiltext{}{jonas.pereira@ufabc.edu.br;  german.lugones@ufabc.edu.br}

\maketitle

\section{Introduction}

Astrophysical observations point to the fact some parts of stars could be solid. For instance, cooling investigations of stars suggest that their outermost regions could be crystalline, at least after a characteristic cooling time \citep{2009ApJ...698.1020B,2015SSRv..191..239P,2018ApJ...860..148C}. Besides, explanation of quasi-periodic oscillations (QPOs) in the tails of giant flares in magnetars also suggest that solid crusts could explain some of the measured frequencies (interpreted as characteristic frequencies of torsional modes) there \citep{1998ApJ...498L..45D,2007MNRAS.374..256S,2017ARA&A..55..261K,2007MNRAS.375..261S,2009CQGra..26o5016S,2008MNRAS.385.2161S,2009MNRAS.396.1441C,2012MNRAS.423..811C,2017MNRAS.464.3101S}. Given that it is unknown the constitution of the innermost regions of neutron stars (see, e.g., \cite{2018arXiv180901116B} and references therein), several classes of stars are possible. For instance, one could have stars with quark or hadronic cores (solid or perfect fluid) and solid hadronic outer phases. However, in the presence of phases with so different natures, surfaces splitting them might hold nontrivial physics, especially in order to guarantee abrupt changes in constitution. Here we investigate that, focusing on general relativistic surface degrees of freedom induced by perturbation on sharp interfaces splitting two very distinct phases. We show that perfect fluids do not induce them upon perturbations but systems with solid parts might. 

Our work can be seen as a generic study of the relevance of general relativistic surface degrees of freedom induced by perturbations in stars, not well discussed in the literature to the best of our knowledge, which could lead to new insights into seismology of neutron stars \citep{2015PhRvD..92f3009K} and their applications. For example, the possibility of surface degrees of freedom induced by perturbations would change boundary conditions for perturbations in hybrid stars, which might influence quantities such as their tidal deformabilities \citep{2008ApJ...677.1216H, 2010PhRvD..81l3016H} and nonradial eigenfrequencies \citep{2003MNRAS.338..389M,Flores:2013yqa}. 
Tidal deformations in neutron stars are already constrained with gravitational wave (GW) observations \citep{2017PhRvL.119p1101A,2018arXiv180511581T}, which could be used to also constrain hybrid stellar models \citep{2018PhRvL.120z1103M,2018ApJ...857...12N,Paschalidis2018}. Gravitational waves from neutron star nonradial oscillations might be detected with the next generation of GW detectors \citep{2017CQGra..34d4001A, 2007PhRvD..75h4038P}. They hold the promise of probing dense matter properties (see e.g., \citep{FloresPRC2017,2017ApJ...837...67C,2018JCAP...08..046V} and references therein), nonlinear couplings \citep{2007PhRvD..75h4038P}, among many other aspects.
 
This work is structured as follows. Sec. \ref{stab-hyb-stars} is devoted to the study of general relativistic surface quantities induced by radial perturbations in (perfect fluid and solid) hybrid stars. In Sec. \ref{consequences}, we derive some important consequences of induced surface degrees of freedom in hybrid stars, especially focused on practical rules for their stability and the difference between their total and gravitational masses. Finally, in Sec. \ref{discussion} we discuss the main points raised. Unless otherwise stated, we work with geometric units and make use of the Schwarzschild coordinates, take the spacetime signature to be $(-,+,+,+)$, and adopt Einstein's summation convention for any two repeated indices.

\section{Radial stability of hybrid stars with surface degrees of freedom}
\label{stab-hyb-stars}

In this work, we focus on unstrained (not solid) background matter at zero temperature and in chemical equilibrium (cold catalyzed matter) which is in different microphysical states at pressures above and below a certain transition pressure $p_{tr}$. We assume that the phase transition at $p_{tr}$ is of first order.  The low and high pressure phases must be in thermodynamic equilibrium at the transition pressure, which means that the corresponding EOSs must satisfy Gibbs conditions at $p_{tr}$ (continuity of  pressure, chemical potential and temperature). However, both phases have in general a different mass-energy density at $p_{tr}$. 

Once a pair of EOSs have been chosen and properly connected through Gibbs conditions, one can solve the hydrostatic equilibrium equations, thus finding the background solution of the hybrid star. For perfect fluids, the hydrostatic equilibrium equations are (TOV equations)
\begin{eqnarray}
\label{tov1}
&&\frac{dp}{dr} = - \frac{\epsilon m}{r^2}\bigg(1 + \frac{p}{\epsilon}\bigg)
	\bigg(1 + \frac{4\pi p r^3}{m}\bigg)\bigg(1 -
	\frac{2m}{r}\bigg)^{-1},
\\ \nonumber \\
\label{tov3}
&&\frac{dm}{dr} = 4 \pi r^2 \epsilon,
\end{eqnarray}
with $p$, $\epsilon$ and $m$, the pressure, energy density and the gravitational mass at the radial distance $r$, respectively. The background spacetime is assumed to be spherically symmetric, i.e.,
\begin{equation}
\label{dsz_tov}
ds^{2}=-e^{ \nu(r)} dt^{2} + e^{ \lambda(r)} dr^{2} + r^{2}(d\theta^{2}+\sin^{2}{\theta}d\phi^{2}),
\end{equation}
where 
\begin{equation}
\label{tov2}
\frac{d\nu}{dr} = - \frac{2}{\epsilon + p} \frac{dp}{dr}
\end{equation}
and
\begin{equation}
e^{\lambda(r)}=\left[1-\frac{2m(r)}{r} \right]^{-1}.
\label{lambda}
\end{equation}
Although the mass-energy density is discontinuous at a phase splitting surface in the hybrid star, other background quantities such as pressure, chemical potential, metric coefficients and extrinsic curvature \citep{2004reto.book.....P} are continuous everywhere.  

The next step is to perturb all phases. For a unstrained background,  solid aspects of matter might or might not raise at this level (because perturbations lead to relative motions of volume elements, a necessary but not sufficient condition for shear forces to appear), and in this work we will consider both possibilities. The perturbed quantities should not be any but should respect Einstein equations and thermodynamics constraints, such as the conservation of the number of baryons and the laws of thermodynamics. In the case of hybrid stars, additional constrains arise. Perturbations on both sides of the interface will be connected now by boundary conditions, which in principle could be set by the microphysics taking place there. Here, we do not solve the associated pulsation equations but just deal with some of the consequences of the boundary conditions chosen to solve them. As we show in this section,  we have to use the Darmois-Israel matching conditions because the perturbed hybrid star is seen as the union of two spacetimes (different from the spacetimes of the background hybrid star) and we want it to fulfill distributional Einstein's equations. 

\subsection{Thin shell formalism in the spherical case}
\label{sec:3a}

When gluing two spacetimes (in our case physically related to a perturbed hybrid star) at a hypersurface $\Sigma$, one must make sure that the resultant one is also a solution to the now distributional Einstein's equations. This is not trivial due to the nonlinearity of the equations. However, as shown by Darmois and Israel, this can be remedied by the presence of an energy momentum tensor $S_{ab}$ at $\Sigma$ \citep{2004reto.book.....P}.

In the spherically symmetric case,  $S_{ab}$ is always defined in terms of a surface energy density $\sigma$ and an isotropic pressure ${\cal P}$ as \citep{2005CQGra..22.4869L}
\begin{equation}
S_{ab}=(\sigma + {\cal P})u_au_b + {\cal P}h_{ab}\label{Sab},
\end{equation}
where $u^a\equiv dy^a/d\tau$ are the velocity components of $\Sigma$ with respect to a local coordinate system $y^a$, $\tau$ is the proper time at $\Sigma$ and $h_{ab}$ is $\Sigma$'s induced metric \citep{2004reto.book.....P}. From Eq. \eqref{Sab} and ordinary tensor calculus, it is clear that $\sigma$ and ${\cal P}$ must be independent of the choice of the local coordinate system $y^a$ and spacetime coordinate systems $x^{\mu}$ (because $S_{ab}$ itself is invariant under changes of $x^{\mu}$) \citep{2004reto.book.....P}. The same ensues for any quantity just dependent on them. 
The symmetry of the problem, the background TOV equations and the already existent perturbation analysis (see the pioneering works  \citet{Wheeler1965,1964PhRvL..12..114C,1964ApJ...140..417C}) naturally suggest the Schwarzschild coordinates, the reason why we choose them here.

In terms of the Schwarzschild coordinates, one generically has that at a phase-splitting surface  $r^{\pm}=R$ \citep{2014PhRvD..90l3011P,2005CQGra..22.4869L} 
\begin{equation}
\sigma = -\frac{1}{4\pi R}\left[\sqrt{e^{-\lambda} +\dot{R}^2 }\right]^+_-
\label{surf_energy_density},
\end{equation}
where $\dot{R}\doteq dR/d\tau$ with $\tau$ the proper time recorded at $R$ and the jump is taken at $R$.
Besides, ${\cal P}$ is given by \citep{2014PhRvD..90l3011P,2005CQGra..22.4869L}
\begin{equation}
{\cal P}=-\frac{\sigma}{2} + \frac{1}{16\pi} \left[\frac{\nu'(e^{-\lambda} +\dot{R}^2)+2 \ddot{R}+\lambda' \dot{R}^2}{ \sqrt[]{e^{-\lambda} + \dot{R}^2}} \right]^+_- ,
\label{surf_tension}
\end{equation}
where the prime operation has been defined as the radial derivative. Actually, $\sigma$ and ${\cal P}$ are not independent, but are related through \citep{2014PhRvD..90l3011P,2005CQGra..22.4869L}
\begin{equation}
d(4\pi R^2\sigma)= -\left({\cal P} - \frac{\Upsilon R}{2}\right)d(4\pi R^2)\label{dot-sigma},
\end{equation}
where
\begin{equation}
\Upsilon\doteq \frac{1}{8\pi R} \left[(\nu'+\lambda')\;\sqrt[]{e^{-\lambda}+\dot{R}^2}\right]^+_-\label{flux}.
\end{equation}

Given that $\sigma$ and ${\cal P}$ have intrinsic general relativistic terms \citep{2014PhRvD..90l3011P}, they are not the usual (laboratory) surface energy density and usual surface tension, respectively. Thus, it seems reasonable to call ${\cal P}$ the ``thin shell surface tension'' and $\sigma$ the ``thin shell surface energy density''.

\subsection{Slow and Rapid phase transitions}
\label{sec:3c}

In order to \textit{calculate} thin shell surface quantities, we need \textit{given} boundary conditions on phase-splitting surfaces. Conversely and generally, if boundary conditions are not specified, then they can be \textit{chosen} such that surface quantities are null, which is equivalent to the continuity of the perturbed extrinsic curvature (see for instance the analysis of \citet{2002PhRvD..66j4002A} regarding nonradial perturbations). Thus, one way or another, the thin shell formalism imposes constraints on hybrid systems. In our approach we fix boundary conditions (they are given ab initio) because they encapsulate relevant physics we want to probe (see below); therefore, surface quantities cannot be chosen at will.

Here we focus on the effects slow and rapid phase conversions (or phase transitions) of perturbed matter have on boundary conditions in hybrid stars \citep{1989A&A...217..137H,2017arXiv170609371P}.
Slow conversions are related to the stretch and squash of volume elements near a phase-splitting surface without their change of nature (conversion timescales much larger than those of perturbations). Rapid conversions are related to a practically immediate conversion of volume elements from one phase to the other and vice-versa in the vicinity of the discontinuity surface upon any perturbation.  

Though physically complicated, the nature of the conversion can be easily summarized mathematically as a junction condition for the Lagrangian change of $r$, $\Delta r\equiv \xi$, and the Lagrangian perturbation of the pressure, $\Delta p$, at a phase-splitting surface. Slow phase conversions verify \citep{2017arXiv170609371P} 
\begin{equation}
[\xi_s]^+_-=0
\label{xislow},
\end{equation}
while rapid phase transitions satisfy the condition \citep{2017arXiv170609371P}
\begin{equation}
[\xi_r]^+_-=\left [\frac{\Delta p}{p'_0}\right]^+_-\label{xirapid},
\end{equation}
where $p_0'\doteq dp_0/dr$, $p_0$ the background pressure. \textit{(From now on we add the subscript ``0'' to background quantities.)} 

For perfect fluids, it is already known that $[\Delta p]^+_-=0$ \citep{2017arXiv170609371P,2014PhRvD..90l3011P}. However, for systems with solid parts, due to the presence of a shear modulus $\tilde{\mu}$ (see, e.g.,  \citet{2008LRR....11...10C} and references therein for some crust models thereof), $\Delta p$ is in general discontinuous at an interface splitting a fluid phase from a solid phase or also two solid phases with different $\tilde{\mu}$s. The physical reason for this result is because for solid systems the total energy momentum tensor $T^{\mu}_{\nu}$ gains (besides its perfect part) an extra component due to shear stresses when perturbations take place \citep{2011PhRvD..84j3006P, 2015PhRvD..92f3009K}. For interfaces splitting fluid and solid phases, for instance, $[\tilde{\mu}]^+_-\neq 0$ in general and that forces $\Delta p$ also to be discontinuous. Let us see more precisely how that happens. One can show from $(T^{\mu}_{r})_{;\mu}=0$ in the spherically symmetric case that at a phase-splitting surface $r=R$
\begin{equation}
    \left[T^r_r\right]^+_-= -\frac{\nu_0'}{2}e^{-\frac{\lambda_0 (R_0)}{2}}\sigma
    \label{continuous-quantity},
\end{equation}
which is the generalization of the continuity of the radial traction ($T^{r}_{\nu}n^{\nu}$) there \citep{2006PhRvD..74d4040G} in the presence of surface degrees of freedom. For example, for slow phase conversions and Gibbs conditions for the background, $[T^r_r]^+_- = \Delta p +\mbox{radial shear terms (proportional to $\tilde{\mu}$)}$. Then, it follows that the presence of discontinuous shear terms would also render $\Delta p$ discontinuous. We elaborate more precisely on that in the next section.

\subsection{Surface degrees of freedom in perturbed hybrid stars}
\label{sec:3b}

In ordinary hybrid stellar models, it is assumed that $\sigma$ and ${\cal P}$ are much smaller than characteristic reference values for equilibrium situations (background), which we assume here to be the case. In this case one can effectively take $\sigma$ and ${\cal P}$ to be null in equilibrium. In such scenario, naturally $\dot{R}=0$ and thus from Eqs. (\ref{lambda}), (\ref{tov2}) and (\ref{tov1}) it only follows that $[m(r)]^+_-=0$ when Gibbs equilibrium conditions are taken into account \citep{1986bhwd.book.....S,2017arXiv170609371P}. From section \ref{sec:3a}, this means that the background extrinsic curvature components are continuous at the phase-splitting surface. However, when perturbations are present, one has that $\nu(r)=\nu_0(r) + \delta \nu(r)$ and $\lambda(r) = \lambda_0(r) + \delta \lambda(r)$, where $\delta$ is the Eulerian operator \citep{1986bhwd.book.....S}, which means that jumps in some physical quantities at a phase-splitting surface might arise when boundary conditions are given from the beginning. Besides, in the presence of perturbations, in the simplest case (assumed in this work) solid aspects to some phases might also appear \citep{2011PhRvD..84j3006P,2015PhRvD..92f3009K}, which could influence surface degrees of freedom too. Since $\dot{R}$ is of the order of the perturbation, $\dot{R}^2$ is null when only first order terms are retained. 
For future reference, from the theory of radial perturbations in neutron stars, it is known that \citep{1964ApJ...140..417C, Wheeler1965,1973grav.book.....M}
\begin{equation}
\delta \lambda= -8\pi r e^{\lambda_0}(p_0+\epsilon_0)\xi= -(\lambda_0'+\nu_0')\xi\label{delta_lambda},
\end{equation}
where we recall that the index ``$0$'' refers to background quantities. For solid systems just in the presence of perturbations (unstrained backgrounds) \citep{2011PhRvD..84j3006P,2015PhRvD..92f3009K}, it is easy to show that Eq. \eqref{delta_lambda} also ensues. (It follows from the $[rt]$ component of the Einstein equations, which does not have any contribution from shear stresses \citep{2011PhRvD..84j3006P,2015PhRvD..92f3009K}.)

We are now in the position of calculating induced thin shell surface degrees of freedom for solid and perfect fluid stars under slow and rapid phase conversions. We start with perfect fluid hybrid stars. In this case, it follows from the previous section that $[\Delta p]^+_-=0$. For slow phase conversions, from Eqs. (\ref{surf_energy_density}), \eqref{xislow} and \eqref{delta_lambda}, we have that [here the phase-splitting surface is at $R=R_0+ \xi_s(R_0,t)$] 
\begin{eqnarray}
\sigma_{s}^{perf}&=& -\lim_{q\rightarrow 0^+}\left[\frac{e^{-\frac{\lambda}{2}}}{4\pi r}\right]^{R_0+\xi_s+q}_{R_0+\xi_s-q}\equiv -\frac{1}{4\pi R}\left[e^{-(\lambda_0+\delta\lambda)/2}\right]^+_- \nonumber\\
&=&\frac{e^{-\frac{\lambda_0(R_0)}{2}}}{8\pi R_0^2}\left[(2-\nu_0'r)\xi_s\right]^+_-=0
\label{energy_density_perturb_slow}
\end{eqnarray}
For the perturbed surface tension ${\cal P}_{s}^{perf}$, from Eqs. \eqref{surf_tension}, \eqref{xislow} and \eqref{delta_lambda}, it follows that 

\begin{widetext}
\begin{eqnarray}
    {\cal P}_{s}^{perf}&=&-\frac{\sigma_s^{perf}}{2} + \frac{1}{16\pi}[e^{-\frac{\lambda}{2}}\nu']^+_-
    =\frac{e^{-\frac{\lambda_0(R_0)}{2}}}{32\pi} \left[ 2\delta \nu' + \left(\nu_0'\right)^2\xi_s+2\xi_s\nu_0''\right ]^+_-
    \nonumber \\ &=&
    \frac{e^{-\frac{\lambda_0(R_0)}{2}}}{8\pi} \left[4\pi e^{\lambda_0}(r\Delta p+2p_0\xi_s) +\frac{\xi_s p_0'\{2(p_0+\epsilon_0)-rp_0' \}}{r(p_0+\epsilon_0)^2} \right]^+_-=0,
    \label{surf-tension-slow-perf}
\end{eqnarray}
\end{widetext}
since $[\Delta p]^+_-=0$ and $\left[p_0'/(p_0+\epsilon_0)\xi_s\right]^+_-$=0 [see Eq. \eqref{tov1}]. In the third equality of the above equation we have used \citep{1964ApJ...140..417C, Wheeler1965,1973grav.book.....M}
\begin{equation}
    \delta \nu'= 8\pi e^{\lambda_0}[r\Delta p -\xi(p_0+\epsilon_0-rp_0')]\label{delta-nu-prime},
\end{equation}
while for the third equality (simplification from the background $[\theta \theta]$ component of Einstein equations)
\begin{equation}
    \nu_0''=8\pi e^{\lambda_0}(3p_0+\epsilon_0-rp_0') + \frac{4p_0'(p_0+\epsilon_0-rp_0')}{r(p_0+\epsilon_0)^2}\label{nu-prime-prime}.
\end{equation}
From Eqs. \eqref{energy_density_perturb_slow} and \eqref{surf-tension-slow-perf}, one sees that perfect fluids under slow phase transitions do not induce thin shell surface degrees of freedom on phase-splitting surfaces at any time.

Now we investigate the case of rapid phase transitions in perfect fluid stars. We recall that for this case the phase-splitting surface in the presence of perturbations is at $R= R_0 + \xi_r-\Delta p/p_0'$ (for details about that, see \citet{2017arXiv170609371P}). From Eqs. (\ref{surf_energy_density}) and \eqref{xirapid}, one has that
\begin{eqnarray}
\sigma_r^{perf}&=& -\frac{1}{4\pi R}[e^{-\lambda/2}]^+_-\nonumber \\
&=&-\frac{e^{-\frac{\lambda_0}{2}}}{8\pi R_0^2}\left[\frac{2(\Delta p- \xi_rp_0') + r\Delta p \lambda_0'+ r\xi_rp_0'\nu_0'}{p_0'} \right]^+_-\nonumber \\ &=&0
\label{energy_density_perturb_rap},
\end{eqnarray}
where use has been made of 
\begin{equation}
    \lambda_0'=8\pi r e^{\lambda_0}(p_0+\epsilon_0)-\nu_0'\label{lambda-prime},
\end{equation}
as well as Eq. \eqref{tov2} and $\left[(p_0+\epsilon_0)/p_0'\right]^{R_0^+}_{R_0^-}=0$.
The perturbed surface tension ${\cal P}_r^{perf}$, with the help of Eqs. \eqref{surf_tension} and \eqref{xislow}, can be written as 

\begin{widetext}
\begin{eqnarray}
    {\cal P}_r^{perf}&=&-\frac{\sigma_r^{perf}}{2} + \frac{1}{16\pi}[e^{-\frac{\lambda}{2}}\nu']^+_-
    =\frac{e^{-\frac{\lambda_0}{2}}}{32\pi} \left[\frac{ 2p_0'\delta \nu' + \Delta p \lambda_0'\nu_0' +\xi_rp_0'(\nu_0')^2 -2\nu_0''(\Delta p  - \xi_r p_0')}{p_0'}\right ]^+_- \nonumber\\&=& \frac{e^{-\frac{\lambda_0}{2}}}{8\pi}\left [\frac{\{ 2(p_0+\epsilon_0)-rp_0'\}(\xi_rp_0'-\Delta p)}{(p_0+\epsilon_0)^2 r}+ \frac{4\pi e^{\lambda_0}\{r  p_0' \Delta p + 2(\xi_rp_0'-\Delta p )p_0 - \Delta p (p_0+\epsilon_0) \}}{p_0'} \right]=0
    \label{surf-tension-rapid}
\end{eqnarray}
\end{widetext}
Therefore, likewise to the slow phase conversion case, no thin shell surface degrees of freedom are induced in perfect fluids when the phase conversion is rapid.

Now we investigate solid hybrid stars. In this case, as commented in the previous section, $[\Delta p]^+_-\neq 0$. Besides, now $\delta \nu'$ changes with respect to its perfect fluid counterpart, Eq. \eqref{delta-nu-prime}, and it is given by [a mere replacement of $\delta p$ by $\delta p + \delta(\mbox{radial shear stresses})$]
\begin{equation}
    \delta \nu'_{sol}= 8\pi e^{\lambda_0}\left[r(\delta p+\delta \Pi^r_r) +  -\xi(p_0+\epsilon_0-rp_0')\right]\label{delta-nu-prime-sol},
\end{equation}
where $\delta \Pi^r_r$ summarizes the contribution to the radial shear stresses in the presence of perturbations (see \citet{2011PhRvD..84j3006P,2015PhRvD..92f3009K} for realizations of them).
For slow phase conversions in solid hybrid stars, due to Eqs. \eqref{xislow} and \eqref{continuous-quantity}, it still follows that $\sigma_s^{sol}=0$ and ${\cal P}^{sol}_s=0$. However, nontrivial results hold for solid hybrid stars under rapid phase conversions. Indeed, for $\sigma^{sol}_r$, we have
\begin{eqnarray}
\sigma_r^{sol}&=& -\frac{1}{4\pi R}[e^{-\lambda/2}]^+_-\nonumber\\
&=&-\frac{e^{-\frac{\lambda_0}{2}}}{8\pi R_0^2}\left[\frac{2(\Delta p- \xi_rp_0') + r\Delta p \lambda_0'+ r\xi_rp_0'\nu_0'}{p_0'} \right]^+_-\nonumber \\
&=& -\frac{p_0+\epsilon_0}{p_0'} e^{\frac{\lambda_0}{2}}\left[ \Delta p \right]^+_-
\label{energy_density_perturb_rap_sol}.
\end{eqnarray}
Before calculating the thin shell surface tension for this case, let us expand Eq. \eqref{continuous-quantity}, since $\sigma^{sol}_r\neq 0$. From our assumption to elastic aspects in hybrid stars, one can  generically write $T^r_r= p_0+ \delta p+ \delta \Pi^r_r$. Thus, after some simple calculations it follows from Eqs. \eqref{tov2}, \eqref{continuous-quantity} and  \eqref{energy_density_perturb_rap_sol} that (recall that $R=R_0+ \xi_r-\Delta p/p_0'$)
\begin{equation}
    [\Delta p + \delta \Pi^r_r]^+_-=0\label{final-continuity-quantity},
\end{equation}
which turns out to be exactly the same as for slow reactions in solid stars (since there $R=R_0+\xi_s$ and $\sigma_s^{sol}=0$). From the above equation, one can clearly see that indeed $[\Delta p]^+_-= -[\delta \Pi^r_r]^+_-\neq 0$ in general. (Equation \eqref{final-continuity-quantity} also makes it clear that the Lagrangian change of the pressure is null for perfect fluids, as already known from other methods \citep{2017arXiv170609371P,2015ApJ...801...19P}.)
Now, for the thin shell surface tension induced by rapid reactions in solid hybrid stars, from Eqs. \eqref{xirapid}, \eqref{delta-nu-prime-sol} and \eqref{final-continuity-quantity}, one obtains that
\begin{equation}
    {\cal P}^{sol}_r=-\frac{\sigma_r^{sol}}{2} + \frac{1}{16\pi}[e^{-\frac{\lambda}{2}}\nu']^+_- = 0 \label{surf-tension-rapid-sol}
\end{equation}
%

The fact that rapid conversions in solid stars induce thin shell surface degrees of freedom could have interesting consequences, for instance the induction of a surface mass (general relativistic energy) on a phase-splitting surface. Since the induced thin shell surface energy density is a local quantity, also is the shell's proper mass (note that for first order quantities, such as associated with surface degrees of freedom, it is irrelevant to use $R$ or $R_0$ for functions multiplying them)
\begin{equation}
M_{shell}\doteq 4\pi R^2\sigma\label{mshell}.
\end{equation}
Total mass measurements at infinity will have to take into account the correction factor $e^{\nu/2}$ (related to the gravitational redshift), as we shall show in Eq. \eqref{m1slow-surface}.

From Eqs. \eqref{surf_energy_density}, (\ref{mshell}) and (\ref{lambda}), it follows that 
\begin{equation}
\frac{M_{shell}}{R}= \left(1-\frac{2m^-}{R} \right)^{\frac{1}{2}}- \left(1-\frac{2m^+}{R} \right)^{\frac{1}{2}}\label{Ms_expanded}.
\end{equation}
Given that $M_{shell}= {\cal O}(\xi)$, when only linear terms in $\xi$ are kept, the above equation can be simplified to $[m]^+_-=M_{shell}e^{-\lambda_0(R_0)/2}$. Thus, from Eqs. (\ref{mshell}) and (\ref{Ms_expanded}),
\begin{equation}
[m]^+_-\equiv [m_1]^+_-=M_{shell}e^{-\lambda_0(R_0)/2}\label{jumpm1},
\end{equation}
where it has been assumed that $m(r)=m_0(r)+m_1(r)$, $m_0$ being the background mass function ($[m_0]^+_-=0$ because the background has been assumed not to have surface degrees of freedom--see Appendix \ref{appendix1}), and $m_1$ its first order correction in the presence of perturbations.

In summary, amongst other effects, the presence of surface degrees of freedom at a surface separating any two perturbed phases implies the discontinuity of the mass in the form given by Eq. \eqref{jumpm1}. We stress that in the thin shell formalism what must be continuous is just the induced metric on a hypersurface $\Sigma$ splitting two spacetimes ($h_{ab}\equiv g_{\mu\nu}\frac{\partial x^{\mu}}{\partial y^a}\frac{\partial x^{\nu}}{\partial y^b}$ and $[h_{ab}]^+_-=0$) \citep{2004reto.book.....P}. (It would be ambiguous in general to demand the continuity of the spacetime metrics on both sides of $\Sigma$ because they are dependent on the choice of the coordinate systems there, $x^{\mu}_{\pm}$. This is clearly not the case for $h_{ab}$.) In this case, $\Sigma$ will have a well-defined geometry. Darmois-Israel conditions automatically take that into account \citep{2004reto.book.....P}.

\subsection{First order corrections to hybrid stars}
\label{sec:3d}

In the case of one-phase stars, it is known that $m_1(r)=-4\pi r^2p_0(r)\xi(r)$ \citep{Wheeler1965}, which means that the total mass contribution due to the perturbations, $m_1(R_{\star})$, where $R_{\star}$ is the stellar background radius, is null since $p_0(R_{\star})=0$. This result means that $\Delta M\equiv m_1(R_{\star})=0$ 
and hence the star's mass as measured by external observers does not change to first order corrections.

When phase-splitting surfaces are present in stars, care should be taken regarding jumps. Assuming that the physics is the same in the phases split by a surface of discontinuity (the perturbation equations in each phase of the star are the same and equal to one-phase systems, see Appendix \ref{appendix1}; only boundary conditions change), Harrison et al.'s one-phase solution \citep{Wheeler1965} holds to $r<R$, that is, 
\begin{equation}
m_1^-(r) = -4\pi r^2 p_0^-(r)\xi^-(r)\label{m1slowminus}.
\end{equation}
For $r>R$, due to the new boundary condition given by Eq. (\ref{jumpm1}),
\begin{equation}
m_1^+(r)=-4\pi r^2 p_0^+(r) \xi^+(r) + M_{shell} e^{\nu_0(R_0)/2}e^{-(\nu_0(r)+\lambda_0(r))/2}\label{m1slow}.
\end{equation}
For further details about the above solutions, see Appendix \ref{appendix1}.
Thus, at the surface of the star, it follows from the above equation that
\begin{eqnarray}
m_1(R_{\star})&\equiv & \Delta M= M_{shell} e^{\nu_0(R_0)/2}\label{m1slow-surface},
\end{eqnarray}
where we have used the fact that $\nu_0(R_{\star})+\lambda_0(R_{\star})=0$, due to the match with the exterior Schwarzschild solution. From Eq. (\ref{m1slow-surface}) one can clearly see that the first order correction to the total mass comes from the phase-splitting surface mass as described by observers at infinity (presence of the term $e^{\nu_0(R_0)/2}$), as it should be. Therefore, the induction of a thin shell surface energy density on a phase-splitting interface implies that first order corrections to the gravitational mass of a hybrid star be non-null.

\section{Some consequences of the induction of surface degrees of freedom in solid stars}
\label{consequences}
Here we focus on some relevant byproducts of inducing thin shell surface degrees of freedom in hybrid stars when boundary conditions to $\xi$ are fixed. From our previous analysis, it immediately applies to rapid phase conversions in solid hybrid stars. 

Let us start by recalling the consequence that the total mass of a cold, catalyzed perfect fluid star with a given number of baryons is automatically extremized when the TOV equations describe its equilibrium state \citep{Wheeler1965,1973grav.book.....M}. Consider a variation with almost null frequency of the central density of a background star. Since the evolution of the system is very slow, it would be basically going from one equilibrium configuration (background) to another one (with varied central density), both fulfilling the TOV equations. Recall that any equilibrium state is characterized by a given energy (mass) and number of baryons, both coming from TOV equations and the microphysics assumed. When first order variations of the total mass are null, exactly the case of cold, catalyzed perfect fluid static stars, the energy (and number of baryons) of the above equilibrium configurations must be the same, i.e., the background mass $M_0$. Hence, $(\partial M_0/\partial \rho_c)|_{\omega=0}=0$, where $\rho_c$ the central density and $\omega$ the frequency of the pulsations. 
As argued in sec. 6.8 of  \citet{1986bhwd.book.....S} (or even in Theorem 18 of \citet{Wheeler1965}), stable one phase stars are associated with the branch $\partial M_0/\partial \rho_c\geq 0$. 

However, Eq. \eqref{m1slow-surface} tells us that when thin shell surface degrees of freedom are present, first order variations to the gravitational mass are not zero anymore. From the reasoning of the above paragraph, it follows now that when eigenfrequencies are close to zero $\partial M_0/\partial \rho_c\neq 0$ and thus, from our findings in the previous section, the usual stability rules should not hold for hybrid solid stars under rapid phase conversions. The point in the $M_0-\rho_c$ plot where $\omega=0$ could either be characterized by $\partial M_0/\partial \rho_c$ positive or negative, depending on solid and perfect fluid aspects. Note that the absence of phase-splitting surfaces, as is the case in one-phase stars, would disallow the induction of thin shell surface degrees of freedom and hence the usual stability rule should ensue, as already known from other methods \citep{2004CQGra..21.1559K}. 

Let us consider now the interpretation of the non-constancy of the stellar mass up to first order perturbations, due to the fact that generically Eq. (\ref{m1slow-surface}) is time-dependent. 
When thin shell surface degrees of freedom are present, in general the total energy of a star should be the sum of the gravitational part ($M=M_0+m_1(R_{\star})=M_0+\Delta M$), the work due to ${\cal P}$ when the thin-shell moves and also a contribution related to the fluxes of momentum through phase-splitting surfaces [see Eq. (\ref{dot-sigma})]. Thus, the change of a first order correction to the gravitational mass of the star would be because now it would not constitute its total energy but it would be part thereof. From the energy channels mentioned above, the total energy up to first order of a hybrid star should be (when local quantities are corrected to observers at infinity by means of the gravitational redshift)
\begin{equation}
    E=M_0+\Delta M + e^{\frac{\nu_0(R_0)}{2}}\int \left({\cal P}- \frac{\Upsilon R}{2} \right)dA\label{total_energy},
\end{equation}
where $A$ is the instantaneous area of the phase-splitting surface. From Eq. \eqref{dot-sigma}, one does see that $dE=0$ in first order of perturbations, and hence $E$ is a constant. We finally stress that we are working with stars at null temperatures, so heat production is not an energy channel to the system, nor particle production due to the conservation of the number of baryons. Therefore, Eq. \eqref{total_energy} would encompass all energy contributions due to surface degrees of freedom.

\section{Discussion and Conclusions}
\label{discussion}

Solid aspects of matter are very different from fluid ones and one would then expect that solid stars should have distinctive phenomenology and physical aspects as well. This might especially be the case for solid hybrid stars because of the possible presence of phase-splitting surfaces. We have focused on general relativistic surface degrees of freedom and have shown that they might be induced in the presence of perturbations when some parts of the star are solid, but not when they are constituted of perfect fluid layers. As one expects, the presence of thin shell surface degrees of freedom might spoil the ordinary rules for the stability of stars because they would be intrinsically associated with imperfect fluids.

Regarding first order corrections to the gravitational mass to external observers, Eq. (\ref{m1slow-surface}), we have shown that they are exactly what one would expect intuitively: the shell's (local) mass corrected by a gravitational redshift factor. We note that this result already takes into account the conservation of the number of baryons in the star since its master equation, Eq. (\ref{m1-eq-wheeler}), automatically does. The gravitational mass is no longer the total energy of the system, which in general also includes the work done by internal forces on the thin-shell (due to ${\cal P}$) and fluxes of momentum due to the shell's motions [see Eq. \eqref{total_energy}], and that is the reason it might vary. Another consequence of a non-null first order correction to the gravitational mass would be the break of the usual condition that $\omega =0$ for $\partial M_0/\partial \rho_c=0$. For slow phase transitions this break should already happen in general because Eq. \eqref{xislow} tacitly assumes that matter around a phase-splitting surface is not catalyzed \citep{2017arXiv170609371P}, and hence one of the conditions for some theorems of \citet{Wheeler1965} is violated.

The possible induction of general relativistic surface degrees of freedom upon perturbations in phase-splitting surfaces might also be relevant for asteroseismology. For instance, it might influence the calculation of tidal deformations in hybrid and binary neutron stars, given that tacit boundary conditions are taken there for the calculation of the Love numbers. It could also influence the calculations of eigenfrequencies of nonradial modes in stars, which are sources of gravitational waves. It might also affect quasi-periodic oscillation predictions based on perturbations when stars have solid parts. This would also be valid for test particles outside the star, since changes in the gravitational mass might perturb their orbits. All of these aspects seem relevant in the era of multi-messenger astronomy and we let them to be investigated more precisely elsewhere.

Summing up, in this work we have shown that surface degrees of freedom could be induced upon perturbations with fixed (and physically motivated) boundary conditions in solid hybrid stars. Their origin is general relativistic and hence bear no direct similarity with laboratory surface energy density and surface tension.
Surface degrees of freedom lead perturbed stars not to conserve their gravitational mass because it is now just part of their total energy.  
Finally, our analysis could also be useful for more generic analyses of perturbations in stars, important for example to characterize sources of gravitational waves as well their structural compositions.

\acknowledgements
J.P.P. acknowledges Nils Andersson for insightful discussions and the financial support given by Funda\c c\~ao de Amparo \`a Pesquisa do Estado de S\~ao Paulo (FAPESP) under grants No. 2015/04174-9 and 2017/21384-2. G.L. is thankful to the Brazilian agencies Conselho Nacional de Desenvolvimento Cient\'{\i}fico e Tecnol\'ogico (CNPq) and FAPESP for financial support. 

\onecolumngrid
\appendix
\section{First order perturbations in the mass function}
\label{appendix1}

In this appendix we precisely deduce Eqs. (\ref{m1slowminus}) and (\ref{m1slow}). 
Before doing so, let us establish the distributional mass equation to be solved. We already know it should be $dm/dr=4\pi T^0_0$, but let us show that consistency of the Darmois-Israel formalism leads exactly to it. In the thin shell formalism, distributions are defined in terms of their proper distances to the phase-splitting surfaces \citep{2004reto.book.....P}, which we denote by $l$. $T^0_0$, which will interest us, is given by \citep{2004reto.book.....P}
\begin{equation}
T^0_0=\epsilon^+\theta(l)+ \epsilon^-\theta(-l) +\sigma\delta(l)\label{t00},
\end{equation}
where $\theta(l)$ is the Heaviside step function and $\delta(l)$ is the Dirac delta function \citep{2004reto.book.....P}. However, it is more appealing to work with coordinate distances given that physical quantities are usually written in terms of them. Noting that $\theta(l)=\theta(r-R)$ and
\begin{equation}
\delta(l)=\delta(e^{\lambda/2}(r-R))=e^{-\lambda(R)/2}\delta(r-R)\label{delta},
\end{equation}
it follows that Eq. (\ref{t00}) can be cast as
\begin{eqnarray}
T^0_0&=&\epsilon^+\theta(r-R)+ \epsilon^-\theta(R-r) +\sigma e^{-\lambda(R)/2}\delta(r-R) \equiv \epsilon^+\theta^+ + \epsilon^-\theta^-+\sigma e^{-\lambda(R)/2}\delta(r-R)
= \epsilon^+\theta^+ + \epsilon^-\theta^- +\frac{[m]^+_-}{4\pi R^2}\delta(r-R)
\label{t00coord},
\end{eqnarray}
where we have used Eqs. (\ref{mshell}) and (\ref{jumpm1}) in the third equality of the above equation.
When $m$ is a distribution, $m=m^+\theta^+ + m^-\theta^-$, and by using $d\theta(r-R)/dr = \delta(r-R)$ \citep{2004reto.book.....P}, we have that
\begin{equation}
\frac{dm}{dr}=\frac{dm^+}{dr} \theta^+ + \frac{dm^-}{dr}  \theta^- + [m]^+_- \delta(r-R).
\label{mdistribution2}
\end{equation}
Using Eq. (\ref{tov3}) for each phase ($dm^{\pm}/dr = 4 \pi r^2 \epsilon^{\pm}$) and comparing with Eq. (\ref{t00coord}) we obtain $dm/dr=4\pi T^0_0$,
exactly as one would expect from the promotion of Einstein's equations to distributions in the spherically symmetric case, thus showing the consistency of the approach.

Now we are set to deduce $m_1$. By working with $T^0_0$ in first order for each phase (see appendix B of \citet{Wheeler1965}) or directly from Theorem 3 of \citet{Wheeler1965}, the differential equation describing $m_1$ is given by
\begin{equation}
\frac{dm_1}{da}+ h(a) \, m_1=T(a),\label{m1-eq-wheeler}
\end{equation}
where
\begin{eqnarray}
h(a)  & \equiv & \frac{1}{r}\left(1-\frac{2m_0}{r} \right)^{-\frac{1}{2}}\mu_0 ;\;\;
T(a)   \equiv   \left[\frac{m_0\mu_0 }{r^2} \left(1-\frac{2m_0}{r} \right)^{-\frac{1}{2}} - 8\pi r p_0 \frac{dr}{da} \right]\Delta r  -4\pi r^2 p_0 \frac{d\Delta r}{da},
\label{T}
\end{eqnarray}
$a$ is the number of baryons within the radius $r$, $\Delta r\equiv \xi$ is a small comoving change of $r$ and $\mu_0 = (p_0+\epsilon_0)/n_0$ is the chemical potential of the background matter with $n_0$ its baryon density. Put that way, all variables depend on $a$, even the radial coordinate, and at the surface of the star $a=A$, being $A$ the total number of baryons. Similarly, $m_1(A)\equiv\Delta M$ is the total mass change due to perturbations. 

Let us work first with the innermost phase, $m_1^-(a)$. The generic solution to Eq.~(\ref{m1-eq-wheeler}) can be split into homogeneous $m_{1h}^-$ and particular $m_{1p}^-$ parts and it is given by
\begin{eqnarray}
m_1^-(a)&=& m_{1h}^-(a)+ m_{1p}^-(a) 
= {\cal C}^-\exp\left(-\int_0^a h(\bar{a})d\bar{a} \right)+ \exp\left(-\int_0^a h(\bar{a})d\bar{a} \right)  \int_0^a \exp \left(\int_0^{\bar{a}} h(\tilde{a})d\tilde{a} \right) T(\bar{a})d\bar{a}
\label{m1minus},
\end{eqnarray}
where ${\cal C}^-$ is an arbitrary integration constant. When the boundary condition $m_1^-(0)=0$ is taken into account, which is a consequence of imposing the regularity condition $m^-(0)=0$, it follows from the above equation that ${\cal C}^-=0$. In what follows we omit the ``$-$'' notation to not overload the equations. From $\exp[...]\,r^2\,p_0d \xi/d\bar{a}= d(\exp[...]\,r^2p_0\xi)/d\bar{a}- \xi d(\exp[....]r^2p_0)/d\bar{a}$, which appears in the last term of Eq. (\ref{m1minus}) when the last term of the second equation of Eq. (\ref{T}) is taken into account, after some simplifications to Eq. (\ref{m1minus}) one arrives at
\begin{eqnarray}
m_{1}^-(a)= m_{1p}^-(a) &=& -4\pi \exp\left(-\int_0^a h(\tilde{a})d\tilde{a} \right) \left[ r^2(\bar{a}) p_0(\bar{a}) \xi(\bar{a}) \exp\left(\int_0^{\bar{a}} h(\tilde{a})d\tilde{a} \right)\right]^a_0 + \exp\left(-\int_0^a h(\bar{a})d\bar{a}  \right)\times \nonumber\\ 
&&\times \int_0^ad\bar{a}\left \{\left[ p'_0+ \frac{(p_0+\rho_0)(m_0+4\pi r^3p_0)e^{\lambda_0}}{r^2}\right]\xi \right. \left. \exp \left(\int_0^{\bar{a}} h(\tilde{a})d\tilde{a} \right)\right\} 
= -4\pi r^2p_0^-\xi^- \label{m1minus_final}, 
\end{eqnarray}
where TOV equation [Eq. (\ref{tov1})] has been taken into account in the third equality, $r(0)=0$ and we have restored the notation for the inner phase (``$-$''). The above equation is exactly Eq.~(\ref{m1slowminus}) and is the same as the one-phase result due to \citet{Wheeler1965}.

Let us now solve Eq. (\ref{m1-eq-wheeler}) in the outer phase in order to obtain $m_{1}^+(a)$. Its general solution is very similar to Eq.~(\ref{m1minus}), with the sole modification that the lower limit of integration to the integrals involved is $a_R$, the background number of baryons at the phase-splitting surface [$r(a_R)=R$]; thus
\begin{eqnarray}
m_1^+(a)&=& m_{1h}^+(a)+ m_{1p}^+(a) 
= {\cal C}^+\exp\left(-\int_{a_R}^a h(\bar{a})d\bar{a} \right) + \exp\left( -\int_{a_R}^a h(\bar{a})d\bar{a} \right) \int_{a_R}^a \exp \left( \int_{a_R}^{\bar{a}} h(\tilde{a})d\tilde{a} \right) T(\bar{a})d\bar{a}\label{m1plus},
\end{eqnarray}
where ${\cal C}^+$ is also an arbitrary constant. It is fixed by the condition that $m_1^+(a_R)=m_1^+$, where $m_1^+$ is the mass immediately above the phase transition surface at $R$ [see Eq.~(\ref{jumpm1})]. From Eq. (\ref{m1plus}), one thus has that ${\cal C}^+=m_1^+$. Similar steps to the $m_1^-$ case and the consideration of the TOV equations leads us to 
\begin{eqnarray}
m_1^+(a) &=& m_1^+\exp\left( -\int_{a_R}^a h(\tilde{a})d\tilde{a} \right) - 4\pi \exp\left( -\int_{a_R}^a h(\tilde{a})d\tilde{a} \right)  \left[ r^2(\bar{a}) p_0(\bar{a}) \xi(\bar{a})\exp\left(\int_{a_R}^{\bar{a}} h(\tilde{a})d\tilde{a} \right)\right]^a_{a_R} \nonumber \\
&=& - 4\pi r^2p_0^+\xi^+ + [m_1]^+_-\exp\left( -\int_{a_R}^a h(\tilde{a})d\tilde{a} \right).
\end{eqnarray}
From Eq. (\ref{jumpm1}), the above equation can be simplified to
\begin{equation}
m_1^+(a)= -4\pi r^2p_0^+\xi^+ + M_{shell} e^{-\lambda_0(a_R)/2}\exp\left(-\int_{a_R}^a h(\tilde{a})d\tilde{a} \right)
\label{m1plushderivative}.
\end{equation}
Since $da/dr= 4\pi r^2e^{\lambda_0/2}n_0$ {(baryon number density taking into account the proper volume)} \citep{Wheeler1965}, $h(a)= e^{\lambda_0/2}\mu_0/r$ [see the first equation of Eq.~(\ref{T})], $\mu_0= (p_0+\rho_0)/n_0$ and $\nu_0'+\lambda_0'=8\pi r e^{\lambda_0}(p_0+\epsilon_0)$ (Einstein's equations), one has that
\begin{equation}
h(a)da= h(r)\frac{da}{dr}dr= \frac{1}{2}\left( \frac{d\lambda_0}{dr} + \frac{d\nu_0}{dr} \right)dr \label{hderivative},
\end{equation}
Finally, when Eq. (\ref{hderivative}) is inserted into Eq. (\ref{m1plushderivative}) and its integral limits are rewritten in terms of $R$ and $r$, one arrives at Eq.~(\ref{m1slow}).

\twocolumngrid

\bibliographystyle{apsrev4-1}
\bibliography{conversions_refs}

\end{document}